\documentclass[twocolumn,showpacs,pra]{revtex4-1}
\usepackage[utf8]{inputenc}
\usepackage[T1]{fontenc}
\usepackage{amssymb}
\usepackage{amsmath}
\usepackage{lmodern}

\usepackage{physics}
\usepackage{graphicx}

\usepackage{siunitx}
\usepackage{mathtools}
\usepackage{booktabs}
\usepackage{microtype}

\usepackage{hyperref}
\usepackage{epstopdf}
\hypersetup{
    linkcolor=blue,
    citecolor=blue,
    urlcolor=blue,
    colorlinks=true,
    pdfauthor={I.Yu. Kostyukov, A.A. Golovanov},
    pdftitle={Field ionization in short and extremely intense laser pulses}
}

\newcommand{\avg}[1]{\left\langle #1 \right\rangle}

\newcommand{\critical}{{\textup{cr}}}
\newcommand{\ionization}{{\textup{i}}}
\newcommand{\atomic}{{\textup{a}}}
\newcommand{\hydrogen}{{\textup{H}}}
\newcommand{\tunnel}{{\textup{TI}}}
\newcommand{\exact}{{\textup{exact}}}
\newcommand{\free}{{\textup{FE}}}
\newcommand{\MA}{{\textup{M}}}
\newcommand{\Keldysh}{{\textup{K}}}
\newcommand{\laser}{{\textup{L}}}
\newcommand{\instantaneous}{{\textup{inst}}}

\newcommand{\electron}{{\textup{e}}}

\DeclareMathOperator{\Ai}{Ai}
\DeclareMathOperator{\Bi}{Bi}

\begin{document}

\title{Field ionization in short and extremely intense laser pulses}

\author{I.\,Yu.~Kostyukov}
\email{kost@appl.sci-nnov.ru}

\author{A.\,A.~Golovanov}
\affiliation{Institute of Applied Physics, Russian Academy of Science, 46 Uljanov
str., 603950 Nizhny Novgorod, Russia}

\begin{abstract}
Modern laser systems are able to generate short and intense laser pulses ionizing matter in the poorly explored barrier-suppression regime.
Field ionization in this regime is studied analytically and numerically.
For analytical studies, both the classical and the quantum approaches are used.
Two approximations to solve the time-dependent Schrödinger equation are proposed: the free electron approximation, in which the atomic potential is neglected, and the motionless approximation, in which only the external field term is considered.
In the motionless approximation, the ionization rate in extremely strong fields is derived.
The approximations are applied to several model potentials and are verified using numeric simulations of the Schrödinger equation.
A simple formula of the ionization rate both for the tunnel and the barrier-suppression regimes is proposed.
The formula can be used, for example, in particle-in-cell codes for simulations of the interaction of extremely intense laser fields with matter.
\end{abstract}


\maketitle

\section{Introduction}

Field ionization is one of the first processes which come into play at ultrahigh-intensity laser--matter interaction.
The peak power of some laser facilities exceeds the \SI{5}{PW} level and will be doubled soon \cite{5pw}.
High intensity laser radiation is generated in the form of very short (less than a hundred femtoseconds) laser pulses, so that atoms and molecules are already ionized at the pulse front.
There are also proposals for secondary radiation sources providing even higher intensities.
For example, an attosecond pulse can be generated at the laser--solid interaction in the relativistic oscillating mirror
regime \cite{Tsakiris2006,Dromey2007}.
The intensity of such attosecond pulses can be even higher than that of the driving PW laser pulse, while the pulse duration is shorter \cite{Gordienko2005}.
At the PW level of the laser intensity, the electric field in the focal spot is several orders of magnitude higher than the characteristic atomic field, $E_{\atomic}=m_{\electron}^{2}e^{5}\hbar^{-4}\approx\SI{5.1e9}{V/cm}$, where $e$ and $m_{\electron}$ are the absolute charge and the mass of an electron, and $\hbar$ is the reduced Planck constant.
This leads to the multiionized states of ions in the plasma being produced at laser--matter interaction.
The ionization-induced mechanisms can play an important role in many high-field phenomena and applications like ionization-induced self-injection in laser--plasma accelerators \cite{Pak2010,McGuffey2010,Clayton2010} or triggering of QED cascades by seed electrons produced at the ionization of high-$Z$ atoms \cite{Tamburini2017,Artemenko2017}.

The regimes of the field ionization in strong electromagnetic field can be roughly classified as follows: the multiphoton ionization regime $E\ll E_{\Keldysh}$, the tunnel ionization (TI) regime $E_{\Keldysh}\ll E\ll E_{\critical}$, and the barrier suppression ionization (BSI) regime $E\gg E_{\critical}$, where $E_{\Keldysh}=\omega_{\laser}(2m_{\electron}I_{\ionization})^{1/2}/e$ is the field threshold associated with the Keldysh parameter $\gamma_{\Keldysh}=\omega_{\laser}(2m_{\electron}I_{\ionization})^{1/2}/(eE)=E_{\Keldysh}/E$, $I_{\ionization}$ is the ionization potential of the atom (ion), $\omega_{\laser}$ is the laser frequency.
The first two regimes are investigated theoretically in detail starting from the milestone paper by Keldysh \cite{Keldysh}. It is generally believed \cite{Popov2004,krainov1998,USP2015} that for short and intense laser pulses the field ionization occurs in the tunnel regime, while the multiphoton ionization is negligible.
The static field tunnel ionization rate (without averaging in time over the laser period) based on the Perelomov--Popov--Terent'ev theory \cite{Perelomov1966-1, Popov2004} is
\begin{align}
&\begin{multlined}
    w_{lm}  = \omega_{a} \kappa^{2} C_{\kappa l}^{2} \cdot 2(2l+1)\left(\frac{2}{F}\right)^{2n^{*}-\abs{m}-1} \\
    \times\frac{(l+\abs{m})!}{2^{\abs{m}}(\abs{m})!(l-\abs{m})!}\exp\left(-\frac{2}{3F}\right),\label{Wpp}
\end{multlined}
\\
&C_{\kappa l}^{2} = \frac{2^{2n^{*}-2}}{n^{*}\Gamma(n^{*}+l+1)\Gamma(n^{*}-l)}, 
\end{align}
where $F=E/\left(\kappa^{3}E_{\atomic}\right)$ is the normalized electric field,
$\kappa^{2}=I_{\ionization}/I_{\hydrogen}$,
$n^{*}=Z/\kappa$ is the effective principal quantum number of the ion,
$Z$ is the ion charge number,
$l$ and $m$ are the orbital and magnetic quantum numbers, respectively, $I_{\hydrogen}=m_{\electron}e^{4}/\left(2\hbar^{2}\right)\approx \SI{13.6}{eV}$ is the ionization potential of hydrogen,
$E_{\atomic}=m_{\electron}^{2}e^{5}\hbar^{-4}\approx\SI{5.1e9}{V/cm}$ is the atomic electric field,
$\omega_{\atomic}=m_{\electron}e^{4}\hbar^{-3}\approx \SI{4.1e16}{c^{-1}}$ is the atomic frequency,
$\Gamma(x)$ is the Gamma function \cite{Abramowitz}.
In the limit $n^{*}\gg1$, formula~(\ref{Wpp}) reduces to the ionization rate given by Ammosov, Delone and Krainov in Ref.~\cite{Ammosov1986}.

\begin{figure}[tb]
    \includegraphics{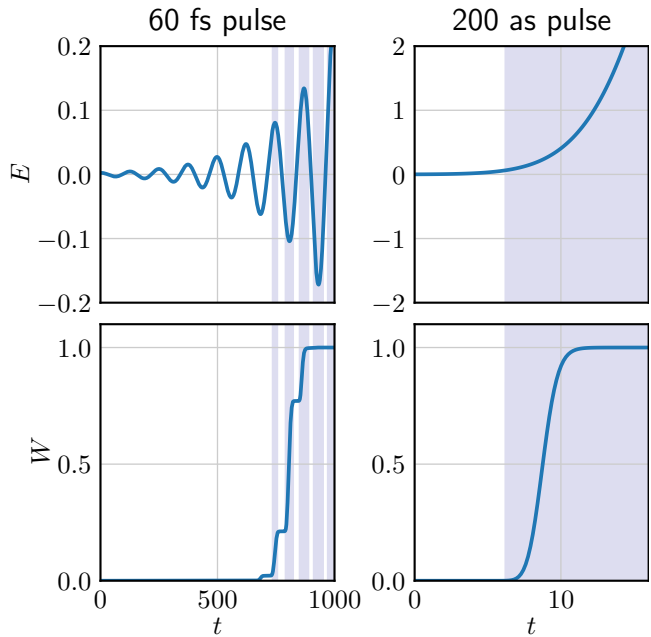}
    \caption{
        Time dependencies of the electric field $E(t)$ and the tunnel ionization probability $W_\ionization(t)$ for hydrogen for a 60 femtosecond Gaussian laser pulse with the wavelength of \SI{0.9}{\um} and the value of $a_0 = eE/(mc\omega_\laser) = 10$ and a 200 attosecond Gaussian video pulse with a maximum field of $10 E_\atomic$.
        The areas where $E > E_\critical$ and the tunnel formula is not applicable are shaded.
        All values are normalized to the atomic units.
    }
    \label{fig:tunnel_applicability}
\end{figure}

The probability of the electron with the minimum ionization potential to leave the atom or ion within the time period $\left[-\infty,\:t\right]$ is equal to 
\begin{equation}
W_{\ionization}(t)=1-\exp\left\{\int_{-\infty}^{t}w\left[E(t')\right]\dd{t'}\right\} ,\label{prob}
\end{equation}
where $w(E)$ is the field ionization rate as a function
of the external field.
The tunnel ionization formulas are no longer valid if the strength of the external field exceeds the atomic critical field, $E_{\critical}=E_{\atomic}\kappa^{4}/\left(16Z\right)$, corresponding to suppression of the atomic potential by the external field.
In this case, the initial energy level of the electron is higher than the maximum of the potential barrier resulted from the superposition of the atomic field and the external field.
In the barrier suppression regime, the electron becomes unbound and propagates above the barrier instead of tunneling.
In real conditions, strong electromagnetic field, $E \gg E_{\critical}$, cannot turn on instantaneously, and there is a finite period of time needed for the laser field to reach the maximum at the atom location.
If such time period is long enough, then the electron with the minimum ionization potential can reach the continuum with the $90\%$ probability at the front of the laser pulse where the tunnel ionization model is valid.
Therefore, the validity of this model for strong laser field depends not only on the field strength, but also on the field rise time.
For simplicity, we assume that the laser field takes a form $E(t)=E_{0}\exp(-4t^{2}/T^{2})$, where the carrier signal
is neglected, $E_{0} > E_\critical$ is the field maximum, and $T$ is the pulse duration. 
In this case, the tunnel ionization model does not break for hydrogen before 90\% ionization is reached if $T \gtrsim \SI{1.6}{ps}$ (see Appendix~\ref{sec:app:tunnel_applicability} for details).
Interestingly enough, the peak field value $E_0$ is not very important in this estimate as long as it exceeds the critical value.

At the PW level of the laser intensity, the field strength exceeds $E_{\critical}$ for the majority of atoms, at least for
the outer electron shells.
The ionization dynamics of the hydrogen atoms in the electromagnetic field is shown in Fig.~\ref{fig:tunnel_applicability}.
The ionization is modeled with tunnel formula (\ref{Wpp}) for two pulses:
(i) for a laser pulse with the Gaussian envelope $E(t)=a_{0}(mc\omega_{\laser}/e)\exp\left(-4t^{2}/T^{2}\right)\cos\left(\omega_{\laser}t\right)$, where $a_{0}=eE_{\laser}/(mc\omega_{\laser})=10$ is the normalized amplitude of the laser field typical for focused sub-PW laser pulses \cite{Poder2018}, $T=60$~fs is the pulse duration, $\lambda_{\laser}=2\pi c/\omega_{\laser}=\SI{0.9}{\um}$ is the laser wavelength;
(ii) for an attosecond pulse with the Gaussian envelope $E(t)=10E_{\atomic}\exp\left(-4t^{2}/T^{2}\right)$, where $T=\SI{200}{as}$.
It is seen from Fig.~\ref{fig:tunnel_applicability} that, even for the laser pulse with the parameters which are typical for existing sub-PW laser systems, the vast majority of the atoms are ionized when $E>E_{\critical}$.
This supports our previous estimate that the tunnel approximation is invalid for sub-ps pulses.

Accurate analytical models of the barrier-suppression regime are absent, since most of the perturbation methods do not work in this regime.
Several formulas for the barrier-suppression ionization rate have been proposed.
The estimate for the barrier-suppression ionization rate based on the classical approach has been derived in Ref.~\cite{Posthumus2018}.
In the limit of infinitely strong external electric field, the rate goes to a constant which does not depend on the strength of the external field.
This result is in a contradiction with the numerical simulations that predict an increase in the ionization rate when increasing the field strength \cite{Bauer1999,Tong2005,Zhang2014}.
The barrier-suppression ionization rate has also been derived in the framework of the Keldysh--Feisal--Reiss theory \cite{Krainov1997}.
However, this model predicts an unphysical decrease in the ionization rate when increasing the field strength in the strong field limit.
Another option for estimating the ionization rate in extremely strong electric field is the empirical approach based on the results of numerical integration of the time-dependent Schrödinger equation (TDSE).
In Ref.~\cite{Bauer1999}, a quadratic rate dependence on the field strength is proposed for $E>E'\sim E_{\critical}$ and the tunnel ionization formula is used for $E<E'$.
The proposed model demonstrates good agreement with the results of numerical simulations for $E\sim E_{\critical}$, but the discrepancy between the model prediction and the numerical results becomes significant in the limit $E \gg E_{\critical}$. There are also difficulties with extending the model beyond hydrogen-like atoms \cite{Artemenko2017}.
Another empirical formula providing continuous transition between the tunnel and the barrier suppression regimes is presented
in Ref.~\cite{Tong2005}.
The model is restricted by the description of ionization of some external shell electrons and of only several chemical
elements.
In the limit $E \gg E_{\critical}$, it predicts unphysical suppression of the ionization rate. 

Numerical simulations are a very powerful and, in some cases, the only tool for the exploration of ultrahigh intensity laser--matter interaction.
Therefore, there is strong demand for a simple formula for the field ionization rate which can be incorporated in particle-in-cell (PIC) codes and can describe a wide range of the electromagnetic field strengths.
One of the simplest numerical models of the strong field ionization is based on the tunnel model for $E<E_{\critical}$, while the electron is assumed unbound if $E\geq E_{\critical}$.
This model significantly overestimates field ionization for $E>E_{\critical}$.
A more accurate formula for the rate assuming linear field dependence for $E>E_{\critical}$ has been proposed in Ref.~\cite{Artemenko2017}.
We will discuss it in more details below.
Numerical models can also include the energy losses associated with ionization \cite{Rae1992, Nuter2011} and can simulate multiple ionization within the time step of the PIC code \cite{Artemenko2017,Nuter2011,Chen2013,Korzhimanov2013}. 

The paper is organized as follows.
In Sec.~\ref{sec:ionization_rate}, the ionization rate formulas in the BSI regime are derived both in the classical and quantum approaches.
The time-dependent Schrödinger equation (TDSE) is analytically integrated in the \emph{free electron approximation}, when the atomic potential is neglected, and in the \emph{motionless approximation}, when the Hamiltonian contains only the external field term.
In Sec.~\ref{sec:models}, the ionization rate is calculated for several model potentials: the 1D $\delta$-potential, the 1D soft-core potential, and the 3D Coulomb potential.
The obtained results are verified by numerical integration of TDSE in Sec.~\ref{sec:numericalSimulations}.
Various formulas for the BSI rate and the validity conditions of the approximations are discussed in Sec.~\ref{sec:conclusions}.

\section{Calculation of ionization rate}
\label{sec:ionization_rate}

In order to estimate the ionization in extremely strong external
electric field, we first use the classical approach \cite{Artemenko2017}.
We assume that (i) the external field is much stronger than the atomic
field at the position of the atomic electron with the ionization
potential $I_{\ionization}$; (ii) the external field turns
on instantaneously: $E=0$ for $t<0$ and $E=\text{const}$
for $t\geq0$; (iii) the electron is ionized at the time instance
$t_{\ionization}$ when it reaches the continuum $\varepsilon_{\textup{free}}=mc^{2}$.
For $t\geq0$, the atomic potential can be neglected, and the electron
will be accelerated in the external field. The initial condition is
$\varepsilon_{0}=mc^{2}-I_{\ionization}$ at $t=0$.
If the atomic forces are neglected, the electron's momentum grows linearly, so that
$\varepsilon(t)=\sqrt{m^2c^4 + (ecEt)^2} - I_\ionization$.
The ionization rate can be estimated as the inverse time $t_{\ionization}$
needed for the electron to reach the continuum $\varepsilon(t_{\ionization})=mc^{2}$
\begin{multline}
    w \approx t_{\ionization}^{-1}=\frac{eE}{\sqrt{2mI_{\ionization}\left(1+\dfrac{I_{\ionization}}{2mc^{2}}\right)}}\approx\frac{eE}{\sqrt{2mI_{\ionization}}}\\
    =\omega_{\atomic}\left(\frac{E}{E_{\atomic}}\right)\sqrt{\frac{I_{\hydrogen}}{I_{\atomic}}},
    \label{eq:rate:ionizationClassical}
\end{multline}
where $(I_{\ionization}/mc^{2})\ll1$ is assumed.
It follows from this estimate that electrons first become ionized before becoming relativistic.
The relativistic corrections
are important only for the inner electrons of high-$Z$ atoms with
very high ionization potentials when the ratio $I_{\ionization}/mc^{2}$
cannot be considered small.

In order to study the BSI regime in the quantum approach, we consider a single-particle nonrelativistic quantum system of an electron in the atomic potential and external uniform varying electric field $\vb{E}(t)$.
We will not consider ionization of electrons with very large $I_{\ionization}\sim mc^{2}$ in order to limit ourselves to the nonrelativistic approximation.
From here and below, we will use atomic units. 
The system is described by a wavefunction $\psi(t, \vb{r})$ and a Hamiltonian \cite{Landau3} 
\begin{equation}
    \hat{H} = \hat{H}_0+ \vb{E}(t) \vb{r}= - \frac{\grad^2}{2} + V(\vb{r}) + \vb{E}(t) \vb{r},
    \label{Hamiltonian}
\end{equation}
where $V(\vb{r})$ is the potential created by the atom.
The influence of the external magnetic field is neglected.
The evolution of the wavefunction satisfies the TDSE
\begin{equation}
    -i \pdv{\psi(t,\vb{r})}{t} = \hat{H} \psi(t,\vb{r}).
\end{equation} 
The static Hamiltonian $\hat{H}_0$ in the absence of the electric field has the bound states $\psi_n(\vb{r})$ with the energies $\epsilon_n$.
In this case, assuming that all wavefunctions are normalized, the probability of the electron to be ionized, i.\,e. to be found in the continuum above all the discrete states, is
\begin{equation}
    W_\ionization(t) = 1 - \sum_{n} \abs{\braket{\psi_n}{\psi(t)}}^2.
    \label{eq:rate:ionizationProbabilityGeneral}
\end{equation}
In general, solving the TDSE analytically is not possible.

However, we consider the case of extremely strong electric fields when the tunnel ionization approximation is invalid.
In this case, we can use the free electron approximation implying that the potential $V(\vb{r})$ is neglected.
Hence, the TDSE in the momentum representation is 
\begin{equation}
    \pdv{\tilde\psi}{t} + \vb{E}(t) \pdv{\tilde\psi}{\vb{p}} = - i \frac{\vb{p}^2}{2} \tilde\psi,
\end{equation}
where 
\begin{equation}
    \tilde\psi(t,\vb{p}) = \frac{1}{(2\pi)^{3/2}} \iiint {\psi(t, \vb{r}) \exp(-i\vb{p}\vb{r}) \dd[3]{\vb{r}}}
\end{equation}
is the wavefunction in the momentum representation.
Using the method of characteristics, it is possible to obtain the solution
\begin{multline}
    \tilde\psi(t, \vb{p}) = \tilde\psi[0, \vb{p} - \vb{A}(t)] \\
    \times \exp\left(-i \int_0^t \frac{[\vb{p} - \vb{A}(t) + \vb{A}(t')]^2}{2} \dd{t'} \right),
    \label{eq:rate:psiEvolutionFreeElectron}
\end{multline}
where $\vb{A}(t) = \int_0^t \vb{E}(t') \dd{t'}$.

In order to determine the probability of the electron to be ionized, Eq.~\eqref{eq:rate:ionizationProbabilityGeneral} can be used.
We assume that initially, at $t=0$, the electron is located in the ground state of $\hat{H}_0$, $\tilde\psi(0,\vb{p}) = \tilde\psi_0(\vb{p})$.
For simplicity, we introduce the quantities
\begin{align}
    &\alpha_n(t) = \braket{\psi_n}{\psi(t)},\quad C_n(t) = \abs{\alpha_n(t)}^2, \\ 
    &C(t) = \sum_n C_n(t).
\end{align}
Here, $C_n(t)$ is the probability of the electron to be found in the $n$th state, and $C(t)$ is the probability of the electron to not be ionized.
At $t=0$, $C_0 = 1$, and all other $C_n = 0$.
Therefore, it is obvious that $C_0(t)$ has the biggest overall contribution to $C(t)$.
The corresponding $\alpha_0$ is calculated as
\begin{multline}
    \alpha_0(t) = \iiint \dd[3]{\vb{p}} \tilde\psi^\ast_0(\vb{p}) \tilde{\psi_0}(\vb{p} - \vb{A}) \\
    \times \exp\left(-i \int_0^t \frac{[\vb{p} - \vb{A}(t) + \vb{A}(t')]^2}{2} \dd{t'} \right).
    \label{eq:rate:alpha0}
\end{multline}
If the field is strong enough and $\vb{A}$ rapidly grows, then the \emph{motionless approximation} can be used in which the exponent in this integral can be neglected (see Sec.~\ref{sec:conclusions} for the details).
This means that the same exponent is neglected in \eqref{eq:rate:psiEvolutionFreeElectron}
\begin{equation}
    \tilde\psi(t, \vb{p}) = \tilde\psi[0, \vb{p} - \vb{A}(t)],
    \label{eq:rate:psiEvolutionMotionless}
\end{equation}
which corresponds to the evolution of a wavefunction described by the Hamiltonian $\hat{H} = \vb{E} \vb{r}$.
In other words, in the total Hamiltonian \eqref{Hamiltonian} not only the atomic potential but also the kinetic energy term $\hat{\vb{p}}^{2}/2=-\grad^{2}/2$ is neglected.
In a bound state, the potential $V(\vb{r})$ and the kinetic energy term $-\grad^{2}/2$ are in balance, so that the squared modulus of the wavefunction remains constant in time, and are typically of the same order.
So the condition for neglecting $V(\vb{r})$ should be the same as for neglecting $-\grad^{2}/2$.
As will be demonstrated by numeric simulations in Sec.~\ref{sec:numericalSimulations}, the motionless approximation is even more accurate than the free electron approximation in the limit of strong external field.

In the coordinate representation, the evolution of the wavefunction is simply phase rotation
\begin{equation}
    \psi(t, \vb{r}) = \psi(0, \vb{r}) \exp[-i \vb{A}(t) \vb{r}].
    \label{eq:rate:psiEvolutionMotionlessCoord}
\end{equation}
This explains our choice of calling this approximation \emph{motionless}, as the electron probability density (equal to $\abs{\psi}^2$) in the coordinate space does not change due to the kinetic energy being neglected.

In the motionless approximation, the probability $C_n(t)$ to find the electron in the $n$th state is
\begin{equation}
    C_n(t) = \abs{\iiint \dd[3]{\vb{p}} \tilde\psi^\ast_n(\vb{p}) \tilde{\psi_0}(\vb{p} - \vb{A}) }^2
\end{equation}
and is determined only by $\vb{A}(t)$.
If we are able to calculate $C(t)=\sum_n C_n(t)$, the instantaneous ionization rate is given by
\begin{equation}
    w_\instantaneous(t) = -\frac{C'(t)}{C(t)}.
\end{equation}
The ionization rate $w_\instantaneous(t)$ inherently depends on $\vb{A}(t)=\int_0^t \vb{E}(t')\dd{t'}$ and thus on the time evolution of the electric field rather than the instantaneous value of the field.
Knowing the initial state of the electron is required as well.
For the use in numerical simulations, such a model might be too complex; also, it cannot describe the transition between the different regimes of ionization.
A simpler model relies on the use of the field ionization rate $w(E)$ which depends on the value of the field.
In this case, the instantaneous ionization rate $w_\instantaneous(t) = w[E(t)]$ depends on the instantaneous strength of the electric field $E(t)$ at the same time moment.
In general, such description is innacurate and can be used as simplification.
In order to estimate $w(E)$, the typical ionization time $t_\ionization = w^{-1}$ can be found in a constant field.
In this case, $C(t) = \tilde{C}(Et)$, and the ionization rate can be estimated by using a transcendental equation
$\tilde{C}(E / w(E)) = \exp(-1)$ whose solution is
\begin{equation}
    w(E) = \frac{E}{\tilde{C}^{-1}\left[\exp(-1)\right]},
    \label{eq:rate:ionizationRate}
\end{equation}
where $\tilde{C}^{-1}$ is the inverse function to $\tilde{C}$.
Similarly to the rate obtained from the classical approach in the beginning of this section, this ionization rate is linear in $E$.
If the ionization process in the constant field is exponential in time, then this formula describes the process exactly.
Otherwise, it serves as an estimate for the ionization rate.
Obviously, there are multiple ways of defining such an estimate.
Other possible estimates for $w(E)$ are discussed in Sec.~\ref{sec:conclusions}.
All of them are linear in $E$ but have different coefficients of the linear dependence.

\section{Models for atomic potential}
\label{sec:models}

\subsection{1D $\delta$-potential}

First, the atomic potential is modeled by the 1D $\delta$-potential.
The Hamiltonian of an electron in this potential in the presence of external electric field is 
\begin{equation}
    \hat{H} = \hat{H}_0 - E x =  -\frac{1}{2}\pdv[2]{x} - \kappa \delta(x) - E(t) x.\label{eq:1d delta}
\end{equation}
If $\kappa$ is positive, $\hat{H}_0$ has only one bound state with the energy of 
\begin{equation}
    \epsilon_0 = - I_0 = - \frac{\kappa^2}{2}.
\end{equation}
If $\kappa = 1$, the energy level $-1/2$ is equal to the energy level of the ground state in a hydrogen atom.
However, unlike the Coulomb potential, there is no critical value of the electric field which completely suppresses the barrier.
The wavefunction of the bound state is 
\begin{align}
    &\psi_0(x) = \sqrt{\kappa} \exp(-\kappa\abs{x}),\\
    &\tilde\psi_0(p) = \sqrt{\frac{2\kappa^3}{\pi}} \frac{1}{\kappa^2 + p^2}
\end{align}
in the coordinate and momentum representations, respectively.

In the motionless approximation \eqref{eq:rate:psiEvolutionMotionless}, the evolution of the wavefunction is described by $\tilde\psi(t, p) = \tilde\psi_0[p - A(t)]$.
As there is only one bound state, the probability $C(t)$ of an electron initially in the bound state to not be ionized is calculated as
\begin{multline}
    C(t) = C_0(t) = \abs{\int_{-\infty}^{\infty} \tilde\psi_0^\ast(p)\tilde\psi_0[p -A(t)] \dd{p}}^2 \\
    = \left[ \qty(\frac{A(t)}{2\kappa})^2 + 1 \right]^{-2}.
    \label{eq:delta1d:C}
\end{multline}

The estimate for the ionization rate according to Eq.~\eqref{eq:rate:ionizationRate} is
\begin{equation}
    w(E) = \frac{E}{2\kappa \sqrt{\sqrt{\exp(1)} - 1}} \approx 0.44 \frac{E}{\sqrt{I_0}}.
\end{equation}
Or, in the physical units,
\begin{equation}
    w(E) = 0.62 \omega_\atomic \frac{E}{E_\atomic} \sqrt{\frac{I_\hydrogen}{I_\ionization}}.
\end{equation}
This formula is almost the same as the classical ionization rate \eqref{eq:rate:ionizationClassical}, which shows that the motionless approximation is conceptually similar to the classical approach in which the electron is accelerated in constant field from the energy of $-I_\ionization$ to $0$.

\subsection{1D soft-core potential}

It is known that the 3D Coulomb potential can be approximated by the so-called soft-core potential \cite{Rae1994} 
\begin{equation}
    V(x) = \frac{Z}{\sqrt{2Z^{-2} + x^2}},
\end{equation}
where $Z$ is the ion charge number.
The properties of the potential $V(x)$ are very similar to the 3D Coulomb potential:
it is long-range, it has an infinite number of bound states, and its ground state energy is equal to $-Z^2/2$.
By the scaling transform, the corresponding Hamiltonian can be reduced to a Hamiltonian with $Z=1$.
Similarly to the Coulomb potential, it has the critical field
\begin{equation}
    E_{\critical} \approx 0.067 Z^3
\end{equation}
above which an electron with the energy of the ground state can pass over the barrier instead of tunneling.
Unlike the 1D $\delta$-potential, the probability of ionization is determined by an infinite sum $C(t) = \sum_n C_n(t)$ as there are multiple eigenstates.
However, the contribution of the eigenstates with a higher number $n$ quickly drops, so taking several lowest-energy functions into account is sufficient.

\begin{figure}[tb]
    \includegraphics{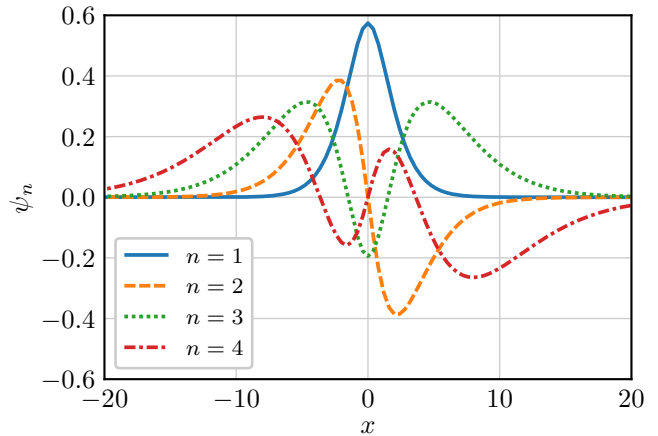}
    \caption{
        Wavefunctions in the soft-core potential with $Z=1$.
        The phase of the wavefunctions is chosen to make the imaginary part equal to zero.
    }
    \label{fig:softcore_wavefunctions}
\end{figure}

The eigenfunctions in this potential have been calculated numerically; the first four eigenfunctions are shown in Fig.~\ref{fig:softcore_wavefunctions}.
Using numerical simulations (see Sec.~\ref{sec:numericalSimulations}), the ionization rate according to Eq.~\eqref{eq:rate:ionizationRate} in the soft-core potential can be estimated as
\begin{equation}
    w (E) \approx 0.87 \frac{E}{Z}.
\end{equation}
In the physical units, it corresponds to 
\begin{equation}
    w(E) \approx 0.87 \omega_\atomic \frac{E}{E_\atomic} \sqrt{\frac{I_\hydrogen}{I_\ionization}}.
\end{equation}
Again, as with the $\delta$-potential, this ionization rate is similar to the classical ionization rate.

\subsection{Coulomb potential}

Some analytical formulas can also be derived for the 3D Coulomb potential 
\begin{equation}
    V(\vb{r}) = -\frac{Z}{r}
\end{equation}
which corresponds to a point-charge ion with the charge $Z$.
By the scaling transform, the problem can always be reduced to $Z = 1$, so only $Z = 1$ is considered from now on.
The bound states in the Coulomb potential are characterized by three quantum numbers $n, l, m$, where $n > 0$, $0 \leq l < n$, $-l \leq m \leq l$.
The energy of a bound state depends only on $n$
\begin{equation}
    \epsilon_n = - \frac{1}{2n^2}.
\end{equation}
The eigenfunctions in the spherical coordinates are \cite{Landau3}
\begin{equation}
    \psi_{n,l,m}(\vb{r}) = R_{n,l}(r) Y_{l,m}(\theta, \varphi),
\end{equation}
where $Y_{l,m}$ are the spherical harmonics, and $R_{n,l}$ are defined as
\begin{equation}
    R_{n,l} = \frac{2}{n^2} \sqrt{\frac{(n-l-1)!}{(n+l)!}} e^{-\frac{r}{n}} \qty(\frac{2r}{n})^l 	L_{n-l-1}^{(2l+1)}\qty(\frac{2r}{n}),
\end{equation}
where $L_n^{(\alpha)}$ are the generalized Laguerre polynomials \cite{Abramowitz}.

It is assumed that uniform electric field is applied to this system, and initially the system is in the ground $\ket{\psi_{1,0,0}}$ state.
Under the motionless approximation, the evolution of the wavefunction can be described by Eq.~\eqref{eq:rate:psiEvolutionMotionlessCoord}.
If we assume that the direction of the electric field always corresponds to the $z$-axis, the evolution of the function in the spherical coordinates is
\begin{equation}
    \psi(t, \vb{r}) = \psi_{1,0,0}(\vb{r}) \exp[-i A(t) r \cos\theta].
    \label{eq:coulomb:psiEvolution}
\end{equation}
The probability of the electron to be found in a bound state is thus
\begin{multline}
    C(t) = \sum_{n,l,m} C_{n,l,m}(t) = \sum_{n,l,m} \Bigg| \int_0^\infty \dd{r} \int_0^\pi \dd\theta \int_0^{2\pi} \dd\varphi  \\
    \times r^2 \sin\theta \psi^\ast_{n,l,m}(\vb{r})\psi_{1,0,0}(\vb{r}) \exp[-i A(t) r \cos\theta]  \Bigg|^2.
\end{multline}
Due to the properties of the spherical harmonics, $C_{n,l,m} \equiv 0$ for $m \neq 0$.
All other integrals can in principle be calculated analytically, as integrands are just polynomials multiplied by an exponent.
For reference, we write down several lowest-order terms,
\begin{align}
    &C_{1,0,0}(t) = \left[1 + \qty(\frac{A(t)}{2})^2\right]^{-4},\\
    &C_{2,0,0}(t) = \frac{8192}{6561} \qty(\frac{2A(t)}{3})^4 \left[1 + \qty(\frac{2A(t)}{3})^2 \right]^{-6}, \\
    &C_{2,1,0}(t) = \frac{8192}{6561} \qty(\frac{2A(t)}{3})^2 \left[1 + \qty(\frac{2A(t)}{3})^2 \right]^{-6}.
\end{align}

\begin{figure}[tb]
    \includegraphics{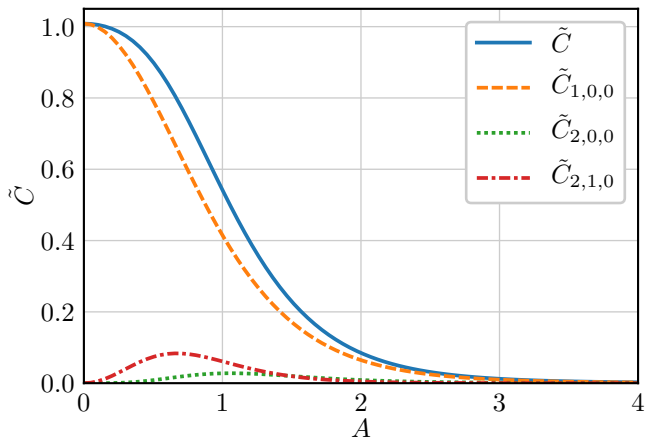}
    \caption{
        Dependencies of $\tilde{C}_{1,0,0}$, $\tilde{C}_{2,0,0}$, $\tilde{C}_{2,1,0}$ and the total sum $\tilde{C} = \sum_{n,l} \tilde{C}_{n,l,0}$ on the integral field $A$ according to model \eqref{eq:coulomb:psiEvolution} for the Coulomb potential with $Z = 1$.
    }
    \label{fig:coulomb_corr}
\end{figure}

The probabilities are fully described by the functions $\tilde{C}_{n,l,m}(A)$ so that $C_{n,l,m}(t) = \tilde{C}_{n,l,m}(A(t))$.
The first three of these functions as well as their total sum are shown in Fig.~\ref{fig:coulomb_corr}.
It is seen that the term $\tilde{C}_{1,0,0}$ is the most dominant factor in $\tilde{C}$, while the influence of the higher-order states is low.
The estimate for the ionization rate in the Coulomb potential given by Eq.~\eqref{eq:rate:ionizationRate} is
\begin{equation}
    w(E) \approx 0.8 \frac{E}{Z}.
\end{equation}
In the physical units, it corresponds to 
\begin{equation}
    w(E) \approx 0.8 \omega_\atomic \frac{E}{E_\atomic} \sqrt{\frac{I_\hydrogen}{I_\ionization}}.
\end{equation}
Once again, like in the considered 1D model potentials, this ionization rate is similar to the classical rate derived in Sec.~\ref{sec:ionization_rate}.
As the numerical constant in hydrogen equal to $0.8$ is close to unity, the classical approach gives a rather accurate description of the ionization rate for extremely strong fields.

\section{Numerical simulations}
\label{sec:numericalSimulations}

\begin{figure}[tb]
    \includegraphics{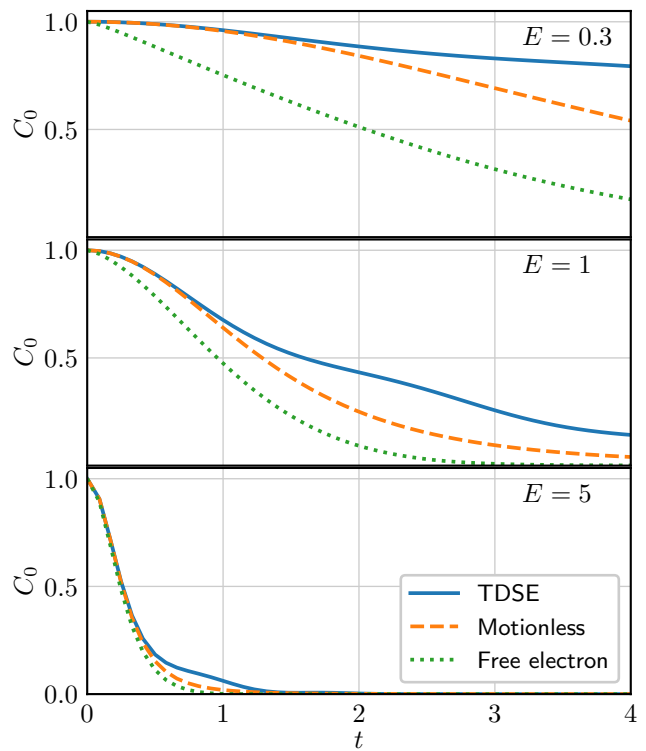}
    \caption{
        Time dependencies of the probability $C_0$ of the electron to be found in the bound state in the $\delta$-potential with $\kappa = 1$ for different values of the electric field $E$ in the numerical integration of the 1D TDSE, in the motionless approximation \eqref{eq:delta1d:C}, and in the free electron approximation \eqref{eq:rate:psiEvolutionFreeElectron}.
    }
    \label{fig:1d_delta_corr}
\end{figure}

In order to demonstrate the applicability of the used approximations, numerical integration of the 1D TDSE was performed for different values of static electric field $E(t)=E$ and the 1D $\delta$-potential with $\kappa=1$.
The Crank--Nicolson method was used for the integration \cite{Crank1947}.
The results of the simulations and their comparison both to the free electron approximation \eqref{eq:rate:psiEvolutionFreeElectron} and the motionless approximation \eqref{eq:delta1d:C} are shown in Fig.~\ref{fig:1d_delta_corr}.
For $\kappa = 1$, our approximations correctly describe the observed behavior for large fields $E > 1$.
For small fields, our models significantly overestimate the ionization rate.
As already mentioned in Sec.~\ref{sec:ionization_rate}, the Hamiltonian $\hat{H}=-E\hat{x}$ corresponding to the motionless approximation even better describes the behavior of the initial system than $\hat{H}=\hat{p}^{2}/2-E\hat{x}$ corresponding to the free electron approximation.
This can also be understood from the fact that, in the limit $t \to 0$, the value of $C_0(t)$ predicted by the motionless approximation is always the same as the exact value of $C_0(t)$ for the total Hamiltonian (see Appendix \ref{sec:app:small_times}).
This behavior is observed in Fig.~\ref{fig:1d_delta_corr}, where the curves for the numerical solution of the TDSE and for the analytical solution in the motionless approximation coincide for small times even for small values of the electric field, when this approximation is not applicable, while the free electron approximation always results in a lower value of $C_0$.

\begin{figure}[tb]
    \includegraphics{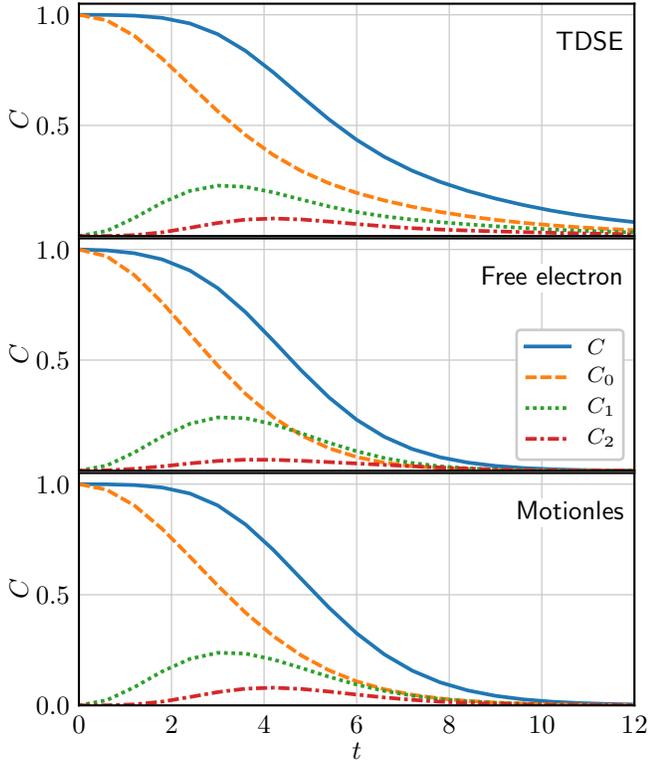}
    \caption{
        Time dependencies of $C_0$, $C_1$, $C_2$, and the total sum $C = \sum_n C_n$ in the numerical integration of the TDSE, the free electron approximation \eqref{eq:rate:psiEvolutionFreeElectron}, and the motionless approximation \eqref{eq:rate:psiEvolutionMotionless} for the soft-core potential with $Z = 1$ and the external electric field strength of $E = 0.2$.
    }
    \label{fig:softcore_multiple}
\end{figure}

The numerical integration of the 1D TDSE was also performed for the soft-core potential with the use of the split-operator spectral solver \cite{Fleck1976, Feit1982}.
Using the ground state as the initial value, the TDSE is integrated for $Z=1$ and various values of stationary electric field.
Figure \ref{fig:softcore_multiple} shows the numerically calculated probabilities $C_{n}(t)$ of the electron to be in the lowest three bound states as well as the total probability $C(t)$ of the electron to be bound for the values of the electric field of $E=0.2>E_{\critical}$.
The corresponding predictions of models \eqref{eq:rate:psiEvolutionFreeElectron} and \eqref{eq:rate:psiEvolutionMotionless} are also shown.
Both the simulations and the models demonstrate similar behavior.
The ground state probability contributes the most to the overall probability of the electron to be in a bound state.
The probabilities of the higher-order states reach their maxima during the process of the electron leaving the atom, but their contribution quickly decreases with the level number.

\begin{figure}[tb]
    \includegraphics{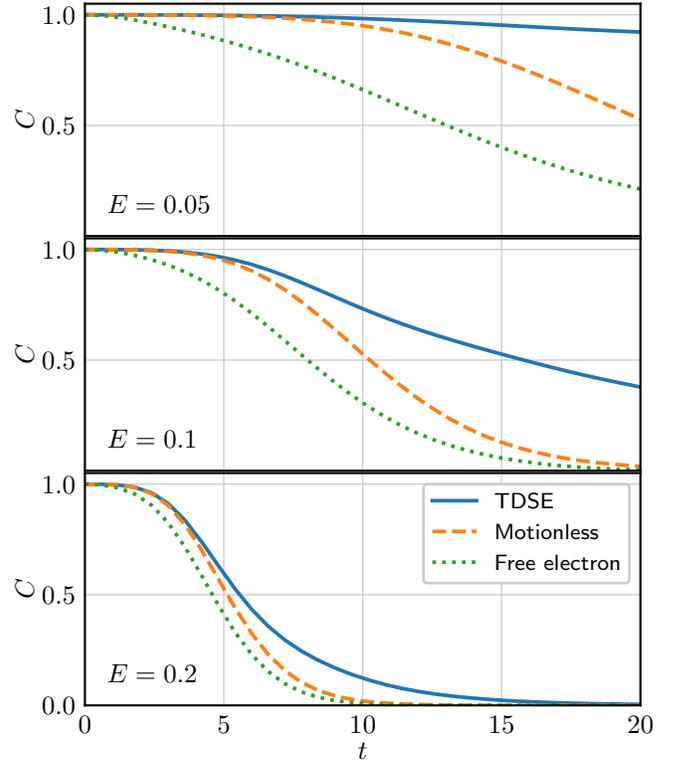}
    \caption{
        Time dependencies of the probability $C(t)$ of the electron to be found in a bound state in the soft-core potential with $Z = 1$ for different values of the electric field $E$ in numerical simulations of the 1D TDSE, in the motionless approximation \eqref{eq:rate:psiEvolutionMotionless}, and in the free electron approximation \eqref{eq:rate:psiEvolutionFreeElectron}.
    }
    \label{fig:softcore_corr}
\end{figure}

Figure \ref{fig:softcore_corr} demonstrates the applicability of our models for different values of stationary electric field for $Z=1$.
For $E=0.05<E_{\critical}$, ionization happens in the tunnel regime, and our approximations are obviously incorrect.
When the external field significantly exceeds the critical field, better correspondence between the simulations and the model is observed.
Again, the Hamiltonian $\hat{H}=-E\hat{x}$ corresponding to the motionless approximation is better suited for the description of the process then $\hat{H}=\hat{p}^{2}/2-E\hat{x}$ corresponding to the free electron approximation.

\begin{figure}[tb]
    \includegraphics{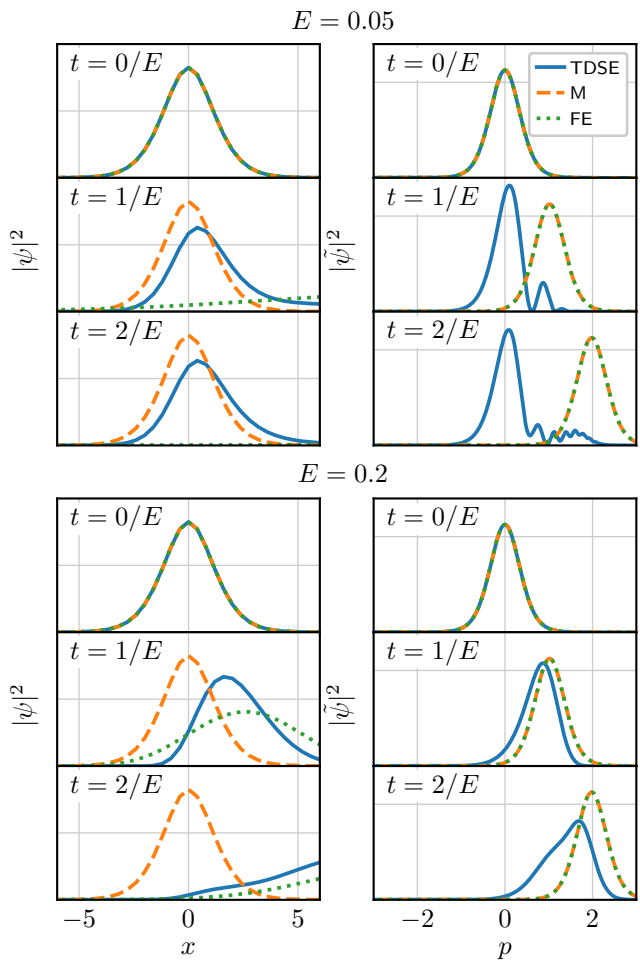}
    \caption{
        Wavefunction probability densities $\abs{\psi}^2$ in the soft-core potential in the coordinate (left column) and the momentum (right column) spaces for different external fields $E$ at different times $t$ in the numerical integration of the TDSE, the motionless (M) approximation, and the free electron (FE) approximation.
    }
    \label{fig:softcore_evolution}
\end{figure}

To better illustrate the correspondence between the approximations and the exact solution according to the TDSE, we plot probability densities $\abs{\psi(x)}^2$ and $\abs{\tilde\psi(p)}^2$ of the wavefunction in the coordinate and the momentum spaces (see Fig.~\ref{fig:softcore_evolution}) at different moments in time for two different values of the electric field: $E = 0.05 < E_\critical$, when the ionization happens in the tunnel regime, and $E = 0.2 > E_\critical$, when the ionization probability is better described by our models, as shown in Fig.~\ref{fig:softcore_corr}.
For both field values, the chosen time moments correspond to the same values of $Et$.
In the momentum space, the probability density is the same in both of our approximations, as the corresponding wavefunctions \eqref{eq:rate:psiEvolutionFreeElectron} and \eqref{eq:rate:psiEvolutionMotionless} differ only in their phases.
Their evolution corresponds to the uniform motion of the probability density.
The exact solution behaves similarly for the overcritical field of $0.2$: the wave packet is accelerated by the electric field while being slightly diffused due to the atomic potential.
As expected, for the tunnel regime in the sub-critical field of $0.05$, our approximations fail, as the probability density in the exact solution remains localized around $p = 0$, and only a tiny fraction of the wave packet is accelerated to higher values of $p$.
In the coordinate space, the two approximations are different.
As the motionless approximation corresponds to the phase rotation in the coordinate space, the corresponding probability density does not evolve at all.
Meanwhile, due to the lack of the potential, the probability density in the free electron approximation dissipates much quicker than in the exact solution, which explains why the values of $C(t)$ in this approximation are always lower (see Fig.~\ref{fig:softcore_corr}).
However, despite the fact that the evolution of the wavefunction is still considerably different in the approximations compared to the exact solution according to the TDSE, the effects taken into account are sufficient to predict the probability of the electron to be ionized.

\section{Discussions and conclusions}
\label{sec:conclusions}

\paragraph{Stark effect in the strong field.}

It is well-known \cite{Popov2004} that there is a relation between
the Stark effect and the field ionization. The Stark width can be
considered as the ionization rate in the limit of the stationary external
field. Therefore, the theory of Stark effect can be used as a qualitative
benchmark for strong field ionization. For the 1D $\delta$-potential
and stationary uniform electric field, the equation
for the quasienergy $\epsilon$ can be derived from Eq.~\eqref{eq:1d delta}
\cite{Popov1993}
\begin{equation}
    \Ai(\xi)\left[\Bi(\xi)+i \Ai(\xi)\right]=\frac{\left(2F\right)^{1/3}}{2\pi},\label{quasi}
\end{equation}
where $\xi=-2\epsilon\left(2F\kappa^{3}\right)^{-2/3}$, $F=E\kappa^{-3}$.
In the weak field limit, the imaginary part of $\epsilon$, or the Stark width,
is reduced to the Keldysh formula for the tunnel ionization rate
\begin{equation}
    \Im[\epsilon(F \to 0)]\sim\left(2F\right)^{-2/3}\exp\left(-\frac{2}{3F}\right).
\end{equation} 
In the strong field limit, the ionization rate is
\begin{equation}
    \Im[\epsilon(F\to \infty)]\sim2^{-5/3}e^{-i\pi/3}\kappa^{2}\left(F\ln F\right)^{2/3}.
\end{equation} 
It is interesting to note that the same expression for $\mathrm{Im}[\epsilon(F\rightarrow\infty)]$
is derived for the Coulomb potential \cite{Benassi1979}, and a similar
expression is derived for the 3D $\delta$-potential $\mathrm{Im}[\epsilon(F\rightarrow\infty)]\sim0.44e^{i\pi/3}\kappa^{2}F^{2/3}$
\cite{Manakov2000}.
Therefore, in the strong-field limit, the dependence of the Stark width on the external electric field is close to linear, similarly to our model.
However, the Stark width
and the Stark level are of the same order of magnitude in the strong
field limit and, strictly speaking, the Stark width cannot be treated
as the ionization rate in this limit.

\paragraph{Conditions of applicability of the motionless approximation.}
In Sec.~\ref{sec:ionization_rate}, we derive the evolution of the wavefunction in the motionless approximation assuming that the exponent in Eq.~\eqref{eq:rate:alpha0} can be neglected, which is equivalent to neglecting the $\hat{\vb{p}}^2/2$ term in the Hamiltonian.
Here, we discuss when this assumption is valid.

As the wavefunction $\tilde{\psi}_{0}$ has the typical width $p_{0}$ in the momentum space, the value of the integral in Eq.~\eqref{eq:rate:alpha0} remains significant only for $\abs{\vb{p}}\lesssim p_{0}$ and only while $\psi_{0}^{\ast}(\vb{p})$ and $\psi_{0}(\vb{p}-\textbf{}\vb{A})$ overlap, so that $A=\abs{\vb{A}}\lesssim p_{0}$.
If the electric field is constant, $\vb{E}=E\vb{x}_{0}$, then the integral is significant for $t\lesssim p_{0}/E$. As both $\vb{p}$ and $\vb{A}$ are bound by $p_{0}$ in the exponent, the maximum value of the expontent may be estimated as $\sim p_{0}^{2}t$.
So, in order to neglect the exponent, 
\begin{equation}
    p_{0}^{2}t\sim\frac{p_{0}^{3}}{E}\ll1
    \label{eq:disc:strongFieldApplicability}
\end{equation}
is required.
If the field is strong enough, $E\gg p_{0}^{3}$, then the phase in the exponent is small, and it does not change the value of the integral much.
This condition of applicability can be generalized for time-dependent fields as well.
For example, for linear in time electric field, the condition is $E(t_{0})\gg p_{0}^{3}$, where $t_{0}$ is the moment of time when $A(t_{0})$ becomes comparable to $p_{0}$.

The typical width of the ground state wavefunction in the momentum space is defined as
\begin{equation}
    p_0^2 = \bra{\psi_0} \hat{p}_x^2 \ket{\psi_0},
\end{equation}
where $x$ is the direction of the electric field.
If the potential spherically symmetric, which is usually the case, this direction does not matter.
In this section, we discuss the meaning of this condition for the potentials we have previously considered.

For the 1D delta potential, $p_0 = \kappa$, where $\kappa$ is the depth of the potential.
In the physical units, taking into account that $I_\ionization = \kappa^2/2$, the condition is
\begin{equation}
    E \gg E_\atomic \qty(\frac{I_\ionization}{I_\hydrogen})^{3/2}.
\end{equation}
It is supported by the numerical simulations in Fig.~\ref{fig:1d_delta_corr} where $E = 1$ serves as the threshold value between the tunnel regime and the motionless approximation regime for $\kappa = 1$.
As the critical field does not exist for this potential, no comparison to it can be made.

For the 1D soft-core potential, the value of $p_0$ is obtained numerically
\begin{equation}
    p_0 \approx 0.39 Z.
\end{equation}
Therefore, the condition in the physical units is
\begin{equation}
    E \gg 0.057E_\atomic Z^3 = 0.057 E_\atomic \qty(\frac{I_\ionization}{I_\hydrogen})^{3/2}.
\end{equation}
Compared to the 1D $\delta$-potential, this condition is much less strict, which is explained by the fact that the barrier in the $\delta$-potential case cannot be completely suppressed no matter how strong the field is, while there is the critical field of $E_\critical \approx 0.067 E_\atomic Z^3$ for the soft-core potential.
The condition above essentially means
\begin{equation}
    E \gg E_\critical,
\end{equation}
which is to be expected.
This conclusion supports the findings of the numerical simulations in Fig.~\ref{fig:softcore_corr} which also show that $E_\critical$ serves as the threshold value between the tunnel and the motionless approximation regimes.

For the 3D Coulomb potential, the value of $p_0$ can be found analytically and is equal to $Z/\sqrt{3}$.
Therefore, condition \eqref{eq:disc:strongFieldApplicability} in the physical units becomes
\begin{equation}
    E \gg \frac{\sqrt{3} Z^3}{9} E_\atomic = \frac{\sqrt{3}}{9} E_\atomic \qty(\frac{I_\ionization}{I_\hydrogen})^{3/2}.
\end{equation}
The critical field for the Coulomb potential is $E_\critical = E_\atomic Z^3/16$, so this condition can also be approximately written as
\begin{equation}
    E \gg 3 E_\critical.
\end{equation}

Overall, for all considered potentials, the scaling of the threshold value (above which the motionless approximation becomes applicable) with the ionization energy is always the same.
So, the general understanding is that the motionless approximation regime can be observed when the field significantly exceeds the critical field.
For time-varying fields, the field must reach such high values before the ionization probability becomes large in order for this approximation to be valid.

\paragraph{Ionization rate in the motionless acceleration regime.}

In Sec.~\ref{sec:ionization_rate}, we introduce the ionization rate $w(E)$ as a function of the instantaneous value of the electric field and come to the conclusion that it is linear in $E$ by estimating the typical ionization time $t_\ionization$.
Here, we discuss the validity of this approach and multiple other ways of determining $w(E)$ in detail.

For simplicity, we consider constant field first.
In the motionless approximation, the dependence of the exact probability $W_\exact$ of the electron to be ionized on time is described by $\tilde{C}(A)$
\begin{equation}
    W_\exact(t) = 1 - \tilde{C}(E t),
\end{equation}
where $\tilde{C}(A)$ is determined by the properties of the quantum system and the initial state.
Meanwhile, for $w(E)$, the ionization probability is exponential in time
\begin{equation}
    W_\ionization(t) = 1 - \exp[-w(E) t].
\end{equation}
As evident from Figs.~\ref{fig:coulomb_corr}, \ref{fig:1d_delta_corr}, \ref{fig:softcore_corr}, the ionization process is not described by an exponent in time even for the constant electric field, much unlike the case of the tunnel regime of ionization.
Therefore, the field ionization rate $w(E)$ cannot accurately describe the actual ionization process.
However, using $w(E)$ instead of solving the TDSE may be very useful for practical applications, e.\,g. for taking ionization into account in particle-in-cell codes.
That is why it is important to find such a dependence $w(E)$ that it describes the actual process in the best possible way.

\begin{table}[tb]
    \centering 
    
    \begin{tabular}{@{}lrrr@{}}
        \toprule
        Method & 1D $\delta$ & 1D soft-core & 3D Coulomb \\ \midrule
        $\exp(-1)$ level & 0.62 & 0.87 & 0.80\\
        Least squares & 0.63 & 0.90 & 0.83\\
        LAD & 0.72 & 1.21 & 1.04\\
        Min. difference & 0.53 & 0.69 & 0.66\\
        \bottomrule
    \end{tabular}
    \caption{Values of the coefficient $\alpha$ in the dependence $w = \alpha E$ for different potentials and different minimization methods.}
    \label{table:alphas}
\end{table}

In order to find such a function, the difference $\Delta W(t) = W_\exact(t) - W_\ionization(t)$ should be minimized according to some criterion.
Naturally, the choice of such criterion is ambiguous.
In Sec.~\ref{sec:ionization_rate}, we propose the following criterion for minimization.
We assume that both $W_\exact(t)$ and $W_\ionization(t)$ reach the value of $\exp(-1)$ at the same moment of time $t_\ionization = w^{-1}$.
According to this criterion,
\begin{equation}
    \tilde{C}\left[\frac{E}{w(E)}\right] = \exp(-1), \quad w(E) = \frac{E}{\tilde{C}^{-1}(\exp(-1))},
\end{equation}
so that the ionization rate is linear in $E$.
Here, $\tilde{C}^{-1}$ is the inverse function to $\tilde{C}$.
We consider three other ways of minimizing $\Delta W(t)$:
\begin{itemize}
    \item minimizing $\int_0^\infty \Delta W^2 \dd{t}$ (the least squares method),
    \item minimizing $\int_0^\infty \abs{\Delta W} \dd{t}$ (the least absolute deviations or LAD method),
    \item minimizing $\max_t\abs{\Delta W}$ (the minimum difference method).
\end{itemize}
In all cases, the ionization rate $w(E)$ turns out linear in $E$,
\begin{equation}
    w(E) = \alpha E,
\end{equation}
which is to be expected considering that $E t$ is a similarity parameter in $W_\exact(t)$.
The coefficient $\alpha$ depends on the considered quantum system and the minimization method.
The values of this coefficient for the 1D $\delta$-potential with $\kappa = 1$, the 1D soft-core potential with $Z=1$, and the 3D Coulomb potential with $Z=1$ are shown in Table~\ref{table:alphas} for all of the considered minimization methods.
In all cases, the coefficients are not significantly different from the coefficient equal to unity obtained from the classical consideration.
For numeric simulations, we propose using the coefficient $0.8$ obtained for the 3D Coulomb potential and for the $\exp(-1)$ level method.

As we have calculated the dependence $w(E) = \alpha E$ assuming the constant field, it is also important to check whether our model is valid for time-varying fields $E(t)$.
If we assume that the direction and the sign of the electric field always remains the same, corresponding to $E(t) > 0$, the exact ionization probability and the probability according to the $w(E)$ model are
\begin{align}
    &W_\exact = 1 - \tilde{C}[A(t)], \\
    &W_\ionization = 1 - \exp[-\alpha A(t)].
\end{align}
We see that both solutions depend on the integral field $A(t)$, which indicates that the typical ionization time will be similar even for varying electric fields.

\begin{figure}[tb]
    \includegraphics[]{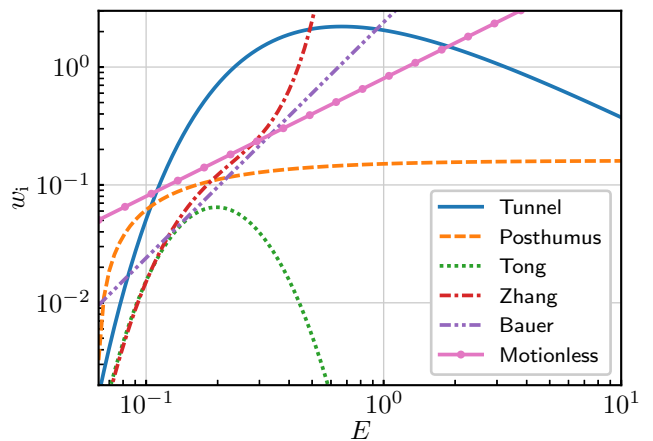}
    \caption{
        Ionization rates for hydrogen according to different models: the tunnel model; the models by Posthumus \emph{et al.} \cite{Posthumus2018}, Tong \emph{et al.} \cite{Tong2005}, Zhang \emph{et al.} \cite{Zhang2014}, Bauer \emph{et al.} \cite{Bauer1999}; and the motionless approximation model proposed in this paper.
        The values are normalized to the atomic units.
    }
    \label{fig:ionization_models}
\end{figure}

\paragraph{Other models of ionization.}
Over the years, several other methods of correcting the tunnel ionization rate have been proposed.
Here, we consider some of them for hydrogen.
For our model, we use the $w = 0.8 E$ formula with the numeric coefficient from Table~\ref{table:alphas}.
Additional formulas include:
\begin{enumerate}
    \item The classical formula by Posthumus \emph{et al.} \cite{Posthumus2018}
    \item The empirical formula by Bauer \emph{et al.} \cite{Bauer1999}
    \item The empirical formula by Tong \emph{et al.} \cite{Tong2005}
    \item The empirical formula by Zhang \emph{et al.} \cite{Zhang2014} with corrections (the minus sign in the exponent in Eq.~(8) of that paper needs to be removed).
\end{enumerate}
These formulas, as well as the motionless approximation regime proposed in this paper, are shown in Fig.~\ref{fig:ionization_models}.

The classical rate by Posthumus \emph{et al.} overestimates the ionization rate at $E \sim E_\critical$ and becomes constant for $E \gg E_\critical$, which makes it a poor estimate.
The three empirical formulas are obtained from numeric integration of the TDSE.
The formula by Bauer \emph{et al.} introduces the empirical scaling $w \propto E^2$ in the area above $E_\critical$.
The formula by Tong \emph{et al.} introduces the empirical correction to the tunnel rate for $E \sim E_\critical$.
For slightly higher fields, the predicted ionization rate quickly drops.
The formula by Zhang \emph{et al.} provides even better empirical approximation, supporting the transition from the model by Tong \emph{et al.} to the model by Bauer \emph{et al.}
However, the drawback of all three models is that they are empirical; they are based on the results of numerical simulations and are applicable only to several chemical elements and several electron shells studied in the corresponding papers.
Unlike that, our formula follows both from classical and quantum analytic considerations and is therefore more general.
It can also be easily applied to different ions if the ground state wavefunction is numerically calculated.

One of the prominent numeric methods of analyzing laser--matter interaction is the particle-in-cell (PIC) method.
Among many other effects, it is possible to take ionization into account in PIC codes.
However, with the amount of different formulas applicable in different ranges, it is important to introduce a formula for the ionization rate which covers most applications.
For example, in the Epoch PIC code \cite{Arber2015}, the formula by Posthumus is used, which---in our opinion---is poorly suitable for the simulations in the BSI regime.
In the Smilei PIC code \cite{Derouillat2018, smilei}, only the tunnel ionization rate is available by default; however, user-defined formulas for the ionization rate may be used instead.
One of the problems arising when making such a choice is that some of the formulas considered above are obtained by fitting the numerical data, which makes it difficult to generalize them for different ions.
A simple approach is to introduce a piecewise formula
\begin{equation}
    w(E) = 
    \begin{cases}
        w_\tunnel(E), &E < E_1,\\
        0.8 \omega_\atomic  \dfrac{E}{E_\atomic} \sqrt{\dfrac{I_\hydrogen}{I_\ionization}}, & E > E_1,
    \end{cases}
\end{equation}
where $w_\tunnel(E)$ is the probability of the tunnel ionization of the specific ion, and $E_1$ is chosen so that $w(E)$ is continuous at $E_1$.
Usually, the linear function crosses the tunnel ionization rate at two points (see Fig.~\ref{fig:ionization_models}), so two possible values of $E_1$ satisfy the continuity condition; the lowest such value has to be chosen.
Except for the numeric coefficient of $0.8$, this ionization rate is the same as proposed in Ref.~\cite{Artemenko2017}.
The numeric coefficient was obtained for the 3D Coulomb potential, so in principle it is also valid only for hydrogen and hydrogen-like ions.
However, even for 1D model potentials, the numeric coefficients are of the same order of magnitude and are reasonably close to unity (see Table~\ref{table:alphas}), so we do not expect major difference for different ions.

To sum up, we have considered field ionization in the limit of extremely strong fields.
Using the classical approach, we have shown that the expected ionization rate is linear with respect to the external field $E$.
To investigate the problem using the quantum approach, we have also considered the single-particle TDSE.
In the strong field limit, two approximations may be used.
One of them is the free electron approximation in which we assume that the atomic potential might be neglected.
The other one is the motionless approximation in which we leave only the term corresponding to the external electric field.
However, the motionless approximation is more accurate in the BSI regime then the free electron approximation.
In the motionless approximation, the ionization rate can be estimated analytically and is always linear in $E$.
For all of the considered model potentials---the 1D $\delta$-potential, the 1D soft-core potential, and the 3D Coulomb potential---the estimated ionization rate is of the same order as predicted by the classical approach.
Numeric integration of the TDSE shows that the motionless approximation is valid when the field strength significantly exceeds the critical value for barrier suppression.
We have also proposed a piecewise formula for the ionization rate both in the tunnel and the BSI regime which can be used, for example, in particle-in-cell codes.

The code used for 1D TDSE integration is available on GitHub \cite{github}.
The Jupyter Notebook producing all figures in the paper is available in the supplementary materials.

\begin{acknowledgments} 

The classical consideration is supported by the Russian Science Foundation through Grant No.~16-12-10383, the quantum consideration and TDSE numeric integration are supported by the Foundation for the advancement of theoretical physics and Mathematics ,,BASIS'' through Grant No.~17-11-101.

\end{acknowledgments}

\appendix
\section{Applicability of the tunnel model}
\label{sec:app:tunnel_applicability}

In order for the tunnel ionization rate $w_\tunnel(E)$ to be valid, most atoms and ions should be ionized before the value of the electric field reaches the critical value $E_\critical$ for the respective orbitals.
In this appendix, we quantitatively evaluate this condition.

For hydrogen, the critical field is $E_\critical = 1/16$, while the ionization rate in the atomic units is 
\begin{equation}
    w_\tunnel(E) = \frac{4}{E} \exp(-\frac{2}{3E}).
\end{equation}
To find the condition of applicability, we assume that the tunnel ionization rate is not applicable if the total probability of ionization \eqref{prob} is less than 90\% when the field reaches the critical value.
This condition corresponds to
\begin{equation}
    \int_{-\infty}^{t_\critical} w_\tunnel[E(t)] \dd{t} < \ln 10,
    \label{eq:app:tunnel_applicability}
\end{equation}
where $E(t_\critical) = E_\critical$.

A very rough estimate may be obtained for arbitrary pulses of full length $T$.
As $w_\tunnel(E) < w_\tunnel(E_\critical)$ for $E < E_\critical$, then if
\begin{equation}
    w_\tunnel(E_\critical) T < \ln 10,
\end{equation}
the tunnel formula is guaranteed to be invalid provided the maximum field strength exceeds the critical value.
In the physical units, it corresponds to 
\begin{equation}
    T < \SI{37}{fs}.
\end{equation}
Therefore, if the pulse is shorter, the tunnel ionization rate can no longer be used.

In a more rigorous approach, a Gaussian video pulse (without the carrier frequency) with $E(t) = E_0 \exp(-4t^2 / T^2)$ is considered.
The value of $E_0$ is assumed to be larger than $E_\critical$.
In this case, the critical field is reached at
\begin{equation}
    t_\critical = - \frac{T}{2} \sqrt{\ln\qty(\frac{E_0}{E_\critical})},
\end{equation}
and condition \eqref{eq:app:tunnel_applicability} corresponds to
\begin{equation}
    T \int_{E_\critical^{-1}}^\infty \frac{1}{\sqrt{\ln(E_0 v)}} \exp(-\frac{2}{3} v) \dd{v} < \ln 10.
\end{equation}
As logarithm is a slow function, if $E_0$ is not too close to $E_\critical$, the approximate value of the left-hand side is
\begin{equation}
    \frac{3T}{2} \qty[\ln(\frac{E_0}{E_\critical})]^{-1/2} \exp(-\frac{2}{3E_\critical}).
\end{equation}
In a wide range of values of $E_0$, the value of the square root of the logarithm can be evaluated as $\sim$1.
Therefore, the tunnel approximation is invalid if
\begin{equation}
    T < \frac{2\ln 10}{3} \exp(\frac{2}{3E_\critical}),
\end{equation}
or, in the physical units,
\begin{equation}
    T < \SI{1.6}{\ps}.
\end{equation}
For pulses with the carrier frequency $\omega_\laser$ and the envelope $E(t)$ the same estimate might be used because
\begin{equation}
    w_\tunnel[E(t) \cos(\omega t)] \leq w_\tunnel[E(t)].
\end{equation}
Therefore, for attosecond and femtosecond pulses with good contrast ratios, the tunnel ionization rate is not valid for hydrogen, and the use of a corrected formula is required.

For hydrogen-like ions with the charge number $Z$, the critical field grows as $Z^3$, and the condition for the pulse duration time becomes 
\begin{equation}
    T < \frac{1.6 \times 10^{-12}}{Z^2}\text{\,s}.
\end{equation}
Therefore, for sufficiently short pulses, the tunnel formula is invalid not only for hydrogen, but for more massive ions as well.

\section{Probability of ionization at small times}
\label{sec:app:small_times}

In this appendix, we explain why the motionless approximation is more accurate than the free electron approximation at small times (see Figs.~\ref{fig:1d_delta_corr}, \ref{fig:softcore_corr}).
We consider three Hamiltonians,
\begin{align}
    &\hat{H}_\exact  = \frac{\hat{\vb{p}}^2}{2} + \hat{V} + \vb{E} \hat{\vb{r}}\\
    &\hat{H}_\free  = \frac{\hat{\vb{p}}^2}{2} + \vb{E} \hat{\vb{r}},\\
    &\hat{H}_\MA  =  \vb{E} \hat{\vb{r}}.
\end{align}
The $\hat{H}_\exact$ Hamiltonian is the initial exact Hamiltonian used in numeric simulations, $\hat{H}_\free$ corresponds to the free electron approximation, and $\hat{H}_\MA$ corresponds to the motionless approximation.
The electric field $\vb{E}$ is assumed to be time-independent.
For each Hamiltonian we calculate
\begin{equation}
    C_{0}^{(i)}(t) = \abs{\braket{\psi_0}{\psi^{(i)}(t)}}^2
\end{equation}
for $t \to 0$.
Here, $\ket{\psi^{(i)}(t)}$ is the solution for $\hat{H}_i$ with the initial condition $\ket{\psi^{(i)}(0)} = \ket{\psi_0}$, $i \in \lbrace \exact, \free, \MA \rbrace$, and $\ket{\psi_0}$ is the eigenfunction of $\hat{H}_0$,
\begin{equation}
    \hat{H}_0 \ket{\psi_0} = H_0 \ket{\psi_0},\quad \hat{H}_0 =  \frac{\hat{\vb{p}}^2}{2} + \hat{V}.
\end{equation}
The evolution of the wavefunctions is described by
\begin{equation}
    \ket{\psi^{(i)}(t)}\approx \left(1 - i \hat{H}_i t - \frac{\hat{H}_i^2 t^2}{2} \right) \ket{\psi_0}
\end{equation}
for $t \to 0$.
Hence, $C_{0}^{(i)}(t)$ in the lowest order in $t$ is 
\begin{equation}
C_{0}^{(i)} \approx 1 - t^2 \left(\avg{\hat{H}_i^2} - \avg{\hat{H}_i}^2\right),
\end{equation}
where $\langle\hat{A}\rangle \equiv \bra{\psi_0}\hat{A} \ket{\psi_0}$.

For $\hat{H}_\exact$, we have
\begin{equation}
\avg{\hat{H}_\exact^2} - \avg{\hat{H}_\exact}^2 = \avg{\qty(\vb{E}\hat{\vb{r}})^2} - \avg{\vb{E} \hat{\vb{r}}}^2.
\end{equation}
For $\hat{H}_\free$, we similarly have
\begin{equation}
\avg{\hat{H}_\free^2} - \avg{\hat{H}_\free}^2 = \avg{\left(\vb{E} \hat{\vb{r}} - \hat{V}\right)^2} - \avg{\vb{E} \hat{\vb{r}} - \hat{V}}^2.
\end{equation}
And for $\hat{H}_\MA$, we have
\begin{equation}
\avg{\hat{H}_\MA^2} - \avg{\hat{H}_\MA}^2 = \avg{\qty(\vb{E} \hat{\vb{r}})^2} - \avg{\vb{E} \hat{\vb{r}}}^2.
\end{equation}
For $\hat{H}_\exact$ and $\hat{H}_\MA$ the answers are identical.
As $C_0(t)$ is the main contributing factor to $C(t)$, this explains why the curves for the numeric solution and the motionless approximation overlap in Figs.~\ref{fig:1d_delta_corr}, \ref{fig:softcore_corr} for small times even when this approximation is not applicable, while the free electron solution is always significantly different.
It may also indicate why the motionless approximation is in general more suitable then the free electron approximation for the description of the system.

\bibliography{Bibliography}

\end{document}